\documentclass[11pt,a4paper]{article}
\usepackage[T1]{fontenc}
\usepackage{newtxtext,newtxmath}
\usepackage[a4paper,margin=2.5cm]{geometry}
\usepackage{microtype}
\usepackage{graphicx}
\graphicspath{{figures/}}
\usepackage{booktabs,tabularx,array,makecell}
\usepackage[font=small,labelfont=bf,labelsep=period]{caption}
\usepackage{enumitem}
\usepackage{placeins}
\usepackage{xcolor}
\usepackage{listings}
\usepackage[numbers,sort&compress,square]{natbib}
\usepackage{url}
\usepackage[hidelinks]{hyperref}

\setlist{nosep,leftmargin=1.5em}


\newcolumntype{L}{>{\raggedright\arraybackslash}X}
\newcolumntype{C}{>{\centering\arraybackslash}X}
\newcolumntype{P}[1]{>{\raggedright\arraybackslash}p{#1}}
\newcolumntype{Q}[1]{>{\centering\arraybackslash}p{#1}}

\newcommand{\armheader}{%
  & & \multicolumn{3}{c}{Rules version} & \multicolumn{3}{c}{Trait-only version} \\
  \cmidrule(lr){3-5}\cmidrule(l){6-8}
  & Neutral & Consultative & Autocratic & Avoidant & Consultative & Autocratic & Avoidant \\}

\title{Consistent but Biased: Chair Personas Trade Noise for Bias in Multi-Agent Clinical Decisions}

\author{%
  \textbf{Yitong Bai}$^{1}$,
  \textbf{Chengyi Zhang}$^{2}$,
  \textbf{Ziyang Wang}$^{3}$\\[0.4em]
  {\small
    $^{1}$Columbia University, USA \quad
    $^{2}$Swansea University, UK \quad
    $^{3}$Aston University, UK
  }
}

\date{}

\begin{document}
\maketitle

\begin{abstract}
\noindent In clinical systems built from several large language model (LLM) agents, one agent often chairs the discussion and makes the final decision. Whether the chair's persona prompt changes that decision, or the agreement between repeated runs that some systems treat as reliability, has not been tested with the case held constant. We ask whether the persona changes the decision or only its timing, how it affects agreement between runs, and which behaviors appear when the prompt describes only personality traits. A simulated department of seven agents discussed 20 published cases of degenerative cervical myelopathy, with the case reports replayed unchanged in every run. Only the prompts of the chair and its deputy changed (no persona, or a consultative, autocratic or avoidant one), written either as rules of conduct or as personality traits. With personas written as rules, the chair chose to operate now in 93\% of runs when autocratic and in 2\% when avoidant, against 57\% without a persona. Surgery was still planned in nearly all decisions, so the personas mainly changed whether to operate now or after more tests. These effects held with a second model and reworded prompts. The autocratic and avoidant personas made the three runs of a case agree more often, but only about as often as chance predicts, and Fleiss $\kappa$ did not rise. Prompts that described only personality traits kept more than half of the autocratic and avoidant effects on operating now. They reproduced quick decisions, delays and extra tests, but not ignoring red flags, dropping backup plans or passing the decision to the family. Studies should report and vary the chair's persona as they do the model, and correct agreement between runs for chance before calling it reliability.

\smallskip
\noindent\textbf{Keywords:} multi-agent systems; persona prompting; clinical decision support; run-to-run consistency; chance-corrected agreement; noise and bias; degenerative cervical myelopathy
\end{abstract}

\section{Introduction}\label{sec:intro}

Surgical decisions depend on who makes them. Rates of lumbar fusion differ nearly 20-fold across regions of the United States \cite{Weinstein2006}, and rates of cervical spine surgery also differ by region \cite{Wang2009}. Spine surgeons given the same lumbar vignettes agree only slightly on management (Fleiss $\kappa$ 0.04--0.13) \cite{Debono2018}, and regions whose surgeons are more enthusiastic about surgery operate more often \cite{Bederman2011}. In multidisciplinary team meetings, the chair runs the discussion and closes it with a plan \cite{Wihl2020}.

Psychology suggests that leaders shape such decisions in three ways. Directive leaders narrow the alternatives and facts a group considers \cite{Leana1985,Flowers1977}. Leaders also change how quickly a decision is closed, since a need for closure pushes toward deciding at once \cite{KruglanskiWebster1996} and choice conflict pushes toward deferral \cite{TverskyShafir1992}. A fixed decision policy, such as one a leader imposes, removes noise, the scatter among judgments that should agree. It also passes any bias in the policy, a systematic shift in one direction, to every decision \cite{Kahneman2021}. In human departments this cannot be tested directly, because the same patient cannot be discussed once under each leader, and studies of leadership in operating rooms and team meetings are observational \cite{Hu2016,Wihl2020}.

Large language model (LLM) multi-agent systems now simulate such meetings. Agents have been organized into multidisciplinary consultations, simulated hospitals and tumor boards \cite{Tang2023,Kim2024,Schmidgall2024,Li2024,Stifini2026}. Many of these systems give one agent the role of moderator or final decision maker, and a short persona prompt sets how that agent behaves. Because their outputs vary from run to run, some clinical systems judge their reliability by whether repeated runs agree \cite{Young2026,ZhangL2026}.

\begin{figure}[htbp]
\centering
\includegraphics[width=\linewidth]{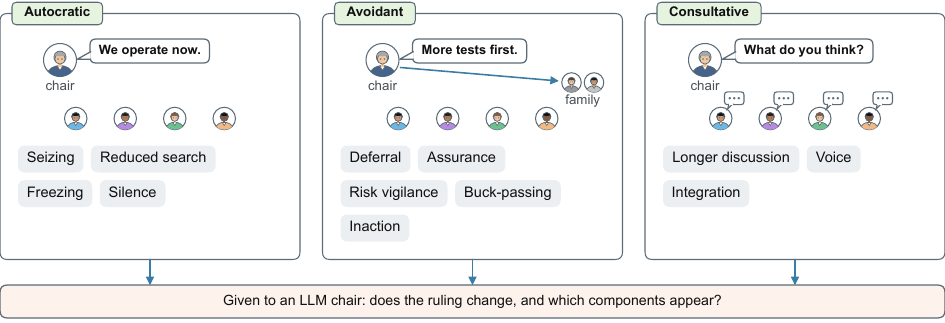}
\caption{Three chair personas and the components that psychology predicts for each (Section~\ref{sec:personas}).}
\label{fig:question}
\end{figure}

A simulated department can discuss the same case under each leader, which human departments cannot, but two questions about such departments are open. First, personas given to LLMs are mostly validated with questionnaire scores, which say little about behavior \cite{Han2025,Contreras2026,Pu2026}, while a leadership construct has several components that a persona may reproduce only in part (Figure~\ref{fig:question}). Second, moderators and leaders change what LLM groups conclude \cite{Reza2025,Kwak2026,Choi2026,Zhu2025}, yet no study has varied the persona of the agent that rules on fixed clinical evidence, or asked whether a persona changes what agreement across runs means (Section~\ref{sec:related}).

We hold the evidence fixed and vary only the persona of the chair and its deputy. A simulated spine surgery department of seven agents discussed 20 published cases of degenerative cervical myelopathy. The statements that report each case and assess its risks were generated once and replayed verbatim in every condition. What varied was the persona of the chair, who alone rules, and of its deputy: no persona (the neutral arm), consultative, autocratic or avoidant. Each persona was written once as rules of conduct and once as traits alone. We ask four research questions (RQs). RQ1: with the evidence fixed, does the chair's persona change the department's decision? RQ2: does it change whether to operate, or only when? RQ3: what does it do to the agreement between repeated runs? RQ4: which components of each persona appear when only its traits are described? The published management is one team's choice, not a reference standard, so we leave the grading of rulings to blinded clinical review and draw conclusions about how such systems are designed and evaluated. Bias here means a systematic shift of an arm's rulings relative to the neutral arm.

The paper makes four contributions:
\begin{itemize}[topsep=3pt,itemsep=2pt]
\item A design for testing the persona of the agent that rules. The case evidence is generated once and replayed, a chair without a persona serves as control, and each persona is written both as rules of conduct and as traits alone, which separates instruction following from persona effects. The design needs only evidence that can be replayed, so it applies to other multi-agent systems.
\item We find that, for the two models tested, the chair's persona moved decisions more than the choice of model. With personas written as rules, the chair ruled to operate now in 2\% (avoidant) to 93\% (autocratic) of runs, against 57\% without a persona. Surgery remained intended in 88--100\% of rulings, so the personas mainly changed when to operate. On a 10-case subset with one run per cell, swapping the decision model moved each arm by 3--20 points.
\item We find that a persona can raise agreement across runs without raising agreement beyond chance (exploratory). The autocratic and avoidant personas raised full agreement across three runs from 50\% to 85--90\% of cases, near the 81--86\% expected by chance from their ruling shares, and Fleiss $\kappa$ did not rise (0.31 without a persona, 0.20 and 0.31 with these personas). Agreement across runs should be reported with its chance level and the distribution of rulings.
\item We find that trait descriptions reproduce part of each persona. Written as traits alone, the autocratic and avoidant personas kept 57\% and 76\% of their effects on operating now. They reproduced quick commitment, deferral and extra testing, but not ignoring red flags, dropping the fallback plan or handing the decision to the family. Alignment training is a candidate explanation, and the meeting's turn order left the silencing of subordinates untested (Section~\ref{sec:limitations}).
\end{itemize}

Section~\ref{sec:related} reviews related work, Section~\ref{sec:methods} describes the department, the personas and the measures, Section~\ref{sec:results} answers the four questions, Section~\ref{sec:limitations} sets out the limits of the design, and Section~\ref{sec:conclusion} concludes.

\section{Related work}\label{sec:related}

\subsection{LLM multi-agent systems for clinical decisions}

LLM agents have been organized into multidisciplinary consultations \cite{Tang2023,Kim2024,Chen2025,Ke2024}, simulated clinics and hospitals \cite{Schmidgall2024,Li2024} and simulated tumor boards \cite{Stifini2026}, and a single LLM has been compared with a breast tumor board \cite{Sorin2023}. Many of these systems give one agent the role of moderator or final decision maker. How that agent is set up changes what the group concludes. A moderator's persona changes the outcome of multi-agent debate \cite{Reza2025}. Leadership styles implemented as controllers, all starting from a shared first round that was generated once and reused, differ in how teams lock in or recover from a wrong initial majority, with little difference in accuracy \cite{Kwak2026}. An agent given authority holds its position while others yield \cite{Choi2026}, and authority labels increase groupthink in LLM collectives \cite{Lin2026}. The structure of the team matters too, with flat teams writing better reports than hierarchical ones \cite{Agachan2026} and elected leaders changing cooperation \cite{Faulkner2026}. In clinical tasks, an interaction style changes the risk behavior of a single model \cite{Abdullahi2026}, and specialist role prompts change GPT-5's language but not its tumor-board decisions \cite{Stifini2026}. An audit of medical multi-agent systems found that authority or majority often stood in for evidence \cite{Zhu2025}, and agents in such systems converge too early unless an agent is added to dissent \cite{WangY2025}. Agreement across repeated runs has been used as a measure of reliability. A clinical decision-support framework was judged reliable because its outputs varied little over five runs \cite{Young2026}, and in a locally deployed diagnostic agent, consistency across runs was the strongest signal of a correct diagnosis \cite{ZhangL2026}. We vary the persona of the agent that rules with the clinical evidence held fixed, and ask what the persona does to such agreement.

\subsection{Personas and personality in LLMs}

LLMs can be given personalities by prompting, and the personalities register on personality questionnaires \cite{SerapioGarcia2023,JiangG2022,JiangH2023}. Personality can also be trained into a model \cite{Maiya2025}. Personas change models' reasoning and risky choices \cite{Gupta2023,Hartley2025}, and LLM agents have been used to simulate human participants \cite{Aher2022,Argyle2023} and human social behavior \cite{Park2023,ZhangJ2023}. Studies that compare self-report with behavior trait by trait find the two dissociated \cite{Han2025,Contreras2026,Pu2026}. Instruction tuning stabilizes how traits are expressed \cite{Han2025}, and shaping personality changes safety-relevant behavior \cite{Fitz2025}. The narrative framing of a task explains several times more behavioral variance than the assigned persona, and the persona effects that carry across narratives come from descriptions whose words map onto actions \cite{WangY2026}. Silence in LLM groups needs a discussion history to anchor it, and a persona alone does not produce it \cite{Zhong2025}. We test, component by component, which parts of a persona appear when it belongs to the agent that rules.

\subsection{Leadership and decision psychology}

Since Lewin's experiments on group climates \cite{Lewin1939}, laboratory tests of Janis's groupthink model \cite{Janis1982} have found that closed or directive leaders lead groups to consider fewer alternatives and facts \cite{Flowers1977,Leana1985}. Need-for-closure theory, conflict theory and research on decision avoidance and defensive medicine describe how decision makers commit, defer or protect themselves \cite{KruglanskiWebster1996,JanisMann1977,Anderson2003,Studdert2005}. Section~\ref{sec:personas} uses these theories to break each persona into components. Error in judgment divides into bias and noise \cite{Kahneman2021}, and surgical judgment is noisy \cite{Irwin2005,Debono2018}. In surgery, leadership style and hierarchy are associated with how much team members speak up \cite{Hu2016,Pattni2019}.

\section{Methods}\label{sec:methods}

\subsection{Cases and pre-decision records}

Cases came from PMC-Patients, a public dataset of 167,034 patient summaries extracted from case reports in PubMed Central \cite{Zhao2023}, used under the CC BY-NC-SA 4.0 license. A keyword search found 545 summaries related to degenerative cervical myelopathy, of which 357 had both a cervical and a myelopathy context. Rules excluded trauma, tumor, infection, inflammatory or demyelinating disease, rheumatoid or atlantoaxial disease, vascular causes and pediatric or congenital causes, and required an age of at least 18 years, leaving 105, and 17 summaries that the rules had wrongly excluded were recovered by hand. On reading each summary, 32 were usable as preoperative cases. After excluding summaries that did not report the actual management, had fewer than about 200 words, focused on surgical technique or spanned a revision operation, we selected 20 cases in three groups: 6 with a borderline indication or no operation, 8 with a difficult differential diagnosis and 6 with a clear indication. According to the full texts, 17 patients had surgery, 2 had surgery after a delay and 1 had no operation. The authors assigned the groups and the actual management from the full texts. We number the cases C1--C20 (Appendix~\ref{app:cases}).

We used the text of each summary but not its title, because titles often state the final management. An LLM (DeepSeek deepseek-v4-pro) deleted the decision and everything after it, cutting within a sentence where needed, and a program verified that each output was a pure deletion of the source (Appendix~\ref{app:redaction}). The redacted records kept a median of 63\% of the original characters (range 24\%--96\%) and had a median length of about 220 words (124--540). No case came with images: the radiologist read the record's description of the imaging and said in its statement that it had not seen the films.

\subsection{The simulated department}\label{sec:department}

\begin{figure}[tb]
\centering
\includegraphics[width=\linewidth]{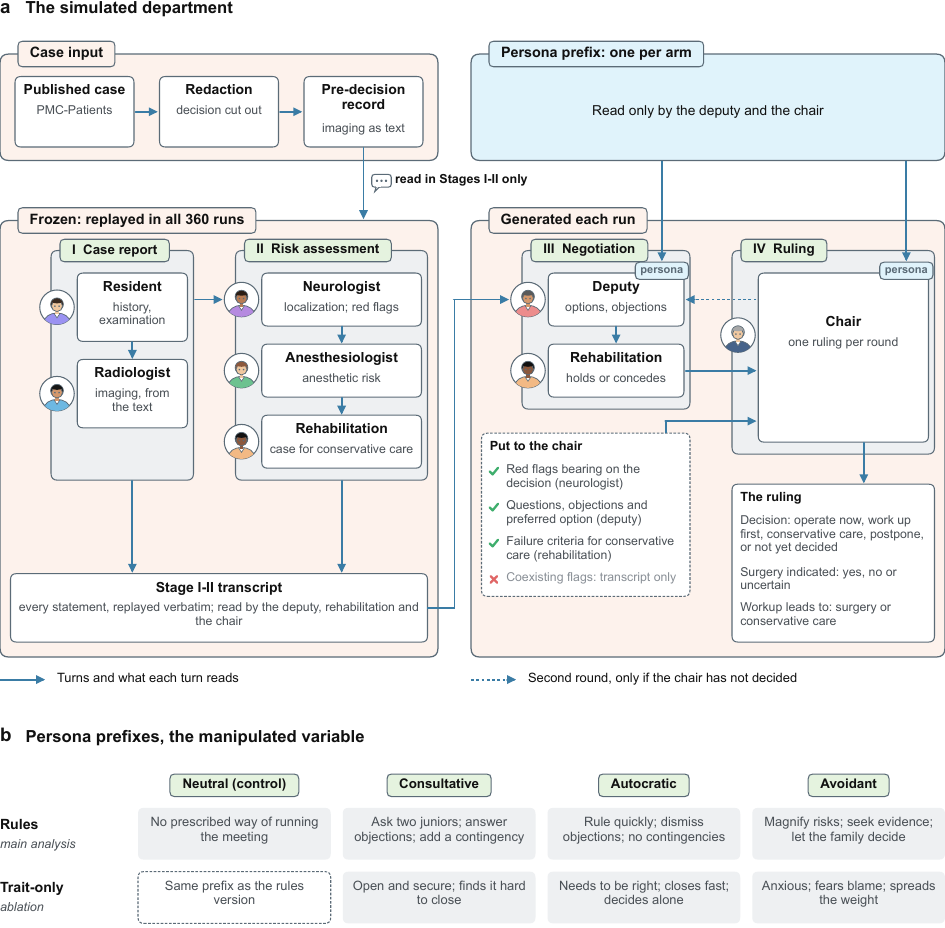}
\caption{The simulated department (a) and the gist of each persona prefix (b), given in full in Appendix~\ref{app:prefixes}.}
\label{fig:system}
\end{figure}

The department has seven roles, and the order of speaking is fixed by the program, not chosen by a model (Figure~\ref{fig:system}). Stage I reports the case. The resident presents the history and examination, and the radiologist reports the imaging findings. Stage II assesses risks. The neurologist localizes the lesion and grades each red flag as changing the indication, changing the plan or coexisting with the surgical decision, and the anesthesiologist assesses anesthetic risk. The rehabilitation physician, as the advocate of conservative care, gives the expected success rate and duration of conservative treatment and the criteria for judging it to have failed. Stage I--II roles read the pre-decision record and the earlier reports they need. From Stage III on, every speaker reads the full transcript. Stage III is negotiation. The deputy speaks first, with options, objections and questions for the chair, and the rehabilitation physician replies, conceding only when someone points to a specific finding already in the record that it had underweighted. Stage IV is the ruling. The chair, prompted as the department director, chooses among operating now, working up first, conservative care, postponing the decision and not yet deciding. It also states whether surgery is indicated (yes, no or uncertain) and, for a workup, whether the workup would lead to surgery or to conservative care. If the chair has not yet decided, Stages III--IV run once more, for at most two rounds. At the ruling, the program also hands the chair a list of items to answer: the neurologist's red flags that bear on the decision, the deputy's questions, objections and preferred option, and the rehabilitation physician's failure criteria for conservative care. Flags graded only as coexisting are not on this list and appear only in the transcript.

Every role returns structured output, and the process measures in this paper, such as the tests the chair orders, its requests of the family and the deputy's objections, are read directly from it. The discussion roles ran on DeepSeek deepseek-v4-pro at its highest reasoning effort and the radiologist on OpenAI gpt-5.6-terra, with the providers' default sampling settings, in October 2026.

\subsection{Persona design}\label{sec:personas}

A \emph{persona} is the prompt prefix that sets how a role conducts itself. An \emph{arm} is one experimental condition: a pair of prefixes, one stating the chair's conduct and one stating the deputy's stance toward that chair. The prompts of all other roles are identical across arms. Each persona exists in two versions. The \emph{rules version}, used in the main analysis, writes conduct as rules. The \emph{trait-only version}, an ablation, describes only traits, the dispositions and motives of the person the role plays. A \emph{component} is a separable psychological process within a persona, such as seizing on a conclusion, that predicts one observable behavior of the department (Figure~\ref{fig:question}; Table~\ref{tab:3} lists each component with its measure).

The autocratic persona draws on closed, directive leadership, under which groups consider fewer alternatives and members comply with a leader's early preference \cite{Leana1985,Flowers1977}. Its psychological bases are a high need for cognitive closure and self-worth that depends on being right \cite{CrockerWolfe2001}. Need for closure comprises an urgency to reach closure (seizing) and a tendency to keep it once reached (freezing), and in groups it breeds pressure toward uniformity and rejection of dissenters \cite{KruglanskiWebster1996,Kruglanski2006}. The persona predicts seizing (more rulings to operate now) and reduced information search (fewer tests). It also predicts freezing, which would leave raised red flags unaddressed and the plan without a fallback, and subordinate silence, a construct distinct from voice \cite{VanDyne2003,Morrison2014}, which would show as fewer objections from the deputy.

The avoidant persona draws on defensive avoidance under decisional conflict, which comprises procrastination, buck-passing and rationalization \cite{JanisMann1977}. Choice conflict leads people, physicians included, to defer \cite{TverskyShafir1992,RedelmeierShafir1995}, and deferring is distinct from not acting \cite{Anderson2003}. Defensive medicine divides into assurance behavior (extra tests and consultations) and avoidance behavior (avoiding high-risk procedures) \cite{Studdert2005,Nahed2012}. The matching temperament is a vigilance to negative stimuli \cite{ElliotThrash2002}, and concern with who will be held responsible, which the avoidant prefixes make salient, changes how judgments are formed \cite{LernerTetlock1999}. The persona predicts deferral (rulings to work up first), possibly inaction (conservative care), buck-passing to the family or to colleagues, assurance behavior (more tests) and risk vigilance (acting on more red flags).

The consultative persona draws on open leadership. Psychological safety and leader openness lead subordinates to speak up \cite{Edmondson1999,Edmondson2003,DetertBurris2007}, and groups under open leaders use more facts \cite{Flowers1977}. Its personal basis is self-worth that does not depend on being right \cite{CrockerWolfe2001}. The persona predicts a chain: a longer discussion, more voice from the deputy, and integration of that voice, so that the ruling changes with what was said.

Two predictions cut across personas. A chair without a persona has no fixed decision policy and should be noisy, whereas a persona that imposes a policy should reduce noise while shifting rulings to one side \cite{Kahneman2021}. Persona effects should also be largest where the indication is ambiguous, because weak situations let individual differences show \cite{Meyer2010,TettBurnett2003}. Mild degenerative cervical myelopathy is such a situation. Guidelines accept both surgery and supervised structured rehabilitation for it \cite{Fehlings2017}, and surgeons in different countries manage it differently \cite{Brannigan2024}.

The neutral chair's prefix says only that no way of running the meeting is prescribed: ``No particular way of running this meeting is prescribed. Run it as you see fit.'' The deputy's neutral prefix likewise does not describe the chair and keeps only the role boundary shared by all four arms: ``You do not rule, and you do not canvass juniors for opinions.'' Appendix~\ref{app:prefixes} gives every prefix in full.

The rules version writes each persona as rules of conduct derived from its components. Before ruling, the consultative chair must ``name at least two junior colleagues and ask them for a specific opinion'', respond formally to reasonable objections and revise the plan where they have merit, and give the plan a contingency. The autocratic chair must rule quickly and reject objections from rehabilitation medicine or junior colleagues ``briskly and in professional terms, for example that the compression is unambiguous and that conservative management would only waste time''. Its plan is short, without elaborate contingencies. The avoidant chair must magnify the risks raised by anesthesia or rehabilitation medicine and seek further evidence, ``move the weight of the decision onto the family wherever you can, for instance by having them sign a detailed informed consent'', and prefer the non-operative option. The deputy's prefix describes which kind of chair it faces and how to behave before it. Before the autocratic chair it offers ``one option, the conservative and safest one you can defend'', compresses its reasoning and blunts its disagreement. Before the consultative chair it raises the contested questions out loud, and before the avoidant chair it puts the question upwards and declines to state a firm preference. Some of the changes in family involvement, deputy objections and contingencies in the rules version were required directly by the prefix and serve as manipulation checks.

The trait-only version describes traits and motives only, taken from the study's original psychological sketches of the three chairs. The autocratic chair reads, among other lines: ``Your self-esteem is closely bound to being the authority who is right. When someone disagrees with you, it can feel less like a point of medicine and more like a challenge to you personally''; ``once you have formed a view, you are inclined to stop weighing further evidence, and you notice what supports your view more readily than what does not''; and ``You see the responsibility for a decision as resting with you alone.'' The avoidant chair reads: ``You are anxious by temperament, and you are drawn to avoiding what you fear''; ``Your fear of an adverse outcome, a complaint or a dispute is stronger than your drive to secure the best possible result''; ``When you weigh a decision, your first thought is who will be held responsible if it goes wrong''; and ``You are uncomfortable carrying the weight of a decision alone, and you are drawn to spreading that weight across as many people as possible.'' The consultative chair is described as open, warm and emotionally steady, secure enough in its own judgment that agreeing that someone else is right does not feel like being proved wrong, and convinced that several perspectives see more than one. Its description ends: ``Under time pressure and a flood of input, you can find it hard to bring things to a close.'' The trait-only prefixes contain no words about treatment, tests, the family, other specialties or the procedure of the meeting (Appendix~\ref{app:prefixes} lists the excluded words). The deputy reads only a description of the chair's traits and decides for itself how to respond. The trait-only version shares the neutral prefixes of the rules version, so the neutral runs serve as the control for both.

To test the role of wording, we also wrote a second rules version for all four arms (version b), with the same rules in different words (Appendix~\ref{app:prefixes} shows the autocratic pair).

\subsection{Holding the evidence fixed}

Stages I--II carry no persona prefix and see none of the negotiation, so they were generated once per case and reused verbatim by every arm and run, which was confirmed for all 360 runs. Differences between arms within a case can come only from the prefixes read by the chair and the deputy and from the model's own randomness, which repeated runs estimate.

Each of the four arms of the rules version ran three times per case (240 runs), and each of the three trait-only arms twice per case (120 runs). One case under one arm is called a cell. Replications used a 10-case subset chosen by rule before any replication result was seen. One replication replaced the model of the chair and the deputy with gpt-5.6-terra, keeping the other roles and the frozen Stage I--II statements, and the other used version b of the rules. Each replication ran once per cell.

\subsection{Measures}\label{sec:measures}

The primary endpoint was the share of runs in which the chair ruled to operate now. Secondary endpoints, specified in advance, were the share of runs in which surgery was judged indicated, the share in which surgery was intended, the share of rulings to work up first, family involvement, the number of objections raised by the deputy, and the handling of red flags. Surgery was counted as intended when the chair ruled to operate now, judged surgery indicated, or recorded in its structured output that the ordered workup would lead to surgery. Appendix~\ref{app:chat} shows how a ruling with two branches is coded. Family involvement meant that the chair put the choice to the family or made the decision wait for a family meeting.

To measure the components we also recorded descriptive measures: the number of tests the chair ordered (information search and assurance behavior), the number of negotiation rounds before the ruling (speed of closure), the number of roles the chair addressed by name in its ruling (spreading the decision across colleagues), whether the plan carried a contingency, and the number of objections the deputy raised in its first turn, before it had heard any ruling. The chair speaks only after the deputy, so subordinate silence can show only as fewer objections in the deputy's own turns.

A judge model (deepseek-v4-pro) coded the handling of red flags. It read the neurologist's list of flags and all of the chair's rulings and coded each flag as acted on (a test or consultation ordered for it, the plan changed because of it, or the plan made conditional on it), mentioned without action (including dismissal) or not addressed. The judge did not know which arm a run belonged to. LLM judges can agree with human preferences as well as humans agree with one another, but they show position, verbosity and other biases \cite{Zheng2023}. A second judge, gpt-5.6-terra, independently coded a random sample of 40 runs (124 flags) and agreed on 91\% of flags (Cohen's $\kappa = 0.84$; Appendix~\ref{app:judge}). \emph{Decision-bearing} red flags are those the neurologist graded as changing the indication or the plan. They were present in 9 cases and are the flags handed to the chair at the ruling.

Noise was defined as the share of cases in which the runs of an arm did not all give the same ruling (three runs in the rules version). For comparison with the trait-only version, which ran twice, we also report the share of cases whose first two runs gave different rulings. By reliability we mean agreement between repeated runs beyond what the arm's ruling shares produce by chance, the sense in which agreement between raters is assessed. Raw agreement rises by itself when one ruling dominates. If one ruling has a 90\% share, three random runs of a case agree 73\% of the time. In an exploratory analysis added after the plan was frozen, we computed the full agreement expected if every run drew its ruling at random with the arm's own ruling shares, and Fleiss' $\kappa$ over the three runs of each case, which corrects agreement for those shares \cite{Fleiss1971}. Both describe repeated runs of one configuration. Neither shows whether another configuration, such as another persona or model, would reach the same rulings.

\section{Results}\label{sec:results}

Table~\ref{tab:hyp} lists the prespecified hypotheses. Each confirmatory test was an exact sign-flip permutation test on the per-case differences from the neutral arm, with Holm correction within each family of tests and a 95\% confidence interval (CI) from a bootstrap over cases. Secondary endpoints used Benjamini--Hochberg (BH) correction (Appendix~\ref{app:stats}). The analysis plan was frozen after the first round of runs and before any repeated run, the trait-only ablation was specified before it was run and the red-flag measure before the flags were coded, and analyses outside the plan are labeled exploratory.

\begin{table}[htbp]
\centering
\footnotesize
\setlength{\tabcolsep}{3.5pt}
\caption{Prespecified hypotheses}
\label{tab:hyp}
\begin{tabularx}{\textwidth}{@{}P{1.0cm}LQ{1.25cm}P{4.6cm}Q{1.0cm}@{}}
\toprule
ID & Hypothesis (against the neutral arm) & Direction & Test & Section \\
\midrule
H1 & Autocratic chair rules to operate now more often & $>$ & Sign-flip, one-sided; Holm over H1--H3 & \ref{sec:rq1} \\
H2 & Avoidant chair rules to operate now less often & $<$ & Sign-flip, one-sided & \ref{sec:rq1} \\
H3 & Consultative chair differs in rulings to operate now & $\neq$ & Sign-flip, two-sided & \ref{sec:rq1} \\
H4 & A persona moves surgery intended less than operating now & -- & Descriptive; bootstrap CI of the gap & \ref{sec:timing} \\
H5 & Share of cases whose three runs agree, per arm & -- & Descriptive & \ref{sec:noise} \\
H6a--c & H1--H3 for the trait-only version & $>$, $<$, $\neq$ & As H1--H3; Holm over H6a--c & \ref{sec:components} \\
H7a & Autocratic chair leaves more decision-bearing flags unaddressed & $>$ & Sign-flip, one-sided; Holm over H7a--c, each version separately & \ref{sec:flags} \\
H7b & Avoidant chair acts on more decision-bearing flags & $>$ & Sign-flip, one-sided & \ref{sec:flags} \\
H7c & Consultative chair differs in decision-bearing flags unaddressed & $\neq$ & Sign-flip, two-sided & \ref{sec:flags} \\
\bottomrule
\end{tabularx}
\end{table}

\subsection{Autocratic and avoidant personas push the decision to opposite extremes (RQ1)}\label{sec:rq1}

\begin{figure}[tb]
\centering
\includegraphics[width=\linewidth]{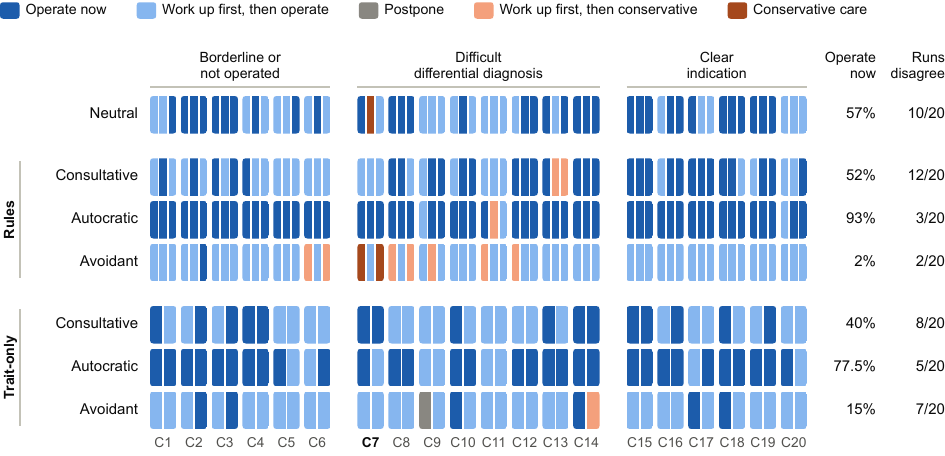}
\caption{Every ruling of the chair (360 runs).}
\label{fig:rulings}
\end{figure}

With the evidence fixed, the chair's persona changed the department's decision (RQ1). The chair ruled within two rounds in all 360 runs. No run of the rules version ended in a postponement or without a decision, and one run of the trait-only avoidant arm postponed. In the rules version, the chair ruled to operate now in 57\% of neutral, 52\% of consultative, 93\% of autocratic and 2\% of avoidant runs (Table~\ref{tab:2}, Figure~\ref{fig:rulings}). The autocratic--neutral difference was +37 percentage points (95\% CI 23 to 50), higher in 13 cases and lower in none (H1, Holm-corrected $p = 0.0002$). The avoidant--neutral difference was $-55$ points (95\% CI $-70$ to $-40$), higher in no case and lower in 17 (H2, Holm $p < 0.0001$). The consultative--neutral difference was $-5$ points (95\% CI $-20$ to 12), higher in 5 cases and lower in 9 (H3, $p = 0.68$).

Four prespecified sensitivity analyses agreed in direction, each ruling out one alternative reading (Appendix~\ref{app:stats}). Modeling single runs instead of each case's share, by logistic regression with a random intercept and by generalized estimating equations, gave the same signs. Counting only each cell's majority ruling, the autocratic persona turned 9 cases toward operating and none away, and the avoidant persona turned 10 away and none toward. Dropping the 11 cases whose redacted records still contained keywords related to the decision, to exclude answers leaked through the redaction, left differences of +33, $-56$ and 0 points. The autocratic and avoidant personas moved every case that had room to move, 13 of 13 and 17 of 17 (Figure~\ref{fig:rulings}), whereas the consultative persona did not move the department detectably. For most cases, the chair's persona decided whether the department operated at once.

\begin{table}[htbp]
\centering
\footnotesize
\setlength{\tabcolsep}{3.5pt}
\caption{Rulings by arm (three runs per case in the rules version, two in the trait-only version)}
\label{tab:2}
\begin{tabularx}{\textwidth}{@{}P{2.9cm}*{7}{C}@{}}
\toprule
\armheader
\midrule
Runs & 60 & 60 & 60 & 60 & 40 & 40 & 40 \\
\addlinespace[3pt]
\multicolumn{8}{@{}l}{\textit{Share of runs}} \\
Operate now & 57\% & 52\% & 93\% & 2\% & 40\% & 77.5\% & 15\% \\
Work up first & 42\% & 48\% & 7\% & 95\% & 60\% & 22.5\% & 82.5\% \\
Conservative care & 2\% & 0 & 0 & 3\% & 0 & 0 & 0 \\
Postpone & 0 & 0 & 0 & 0 & 0 & 0 & 2.5\% \\
Surgery intended & 100\% & 100\% & 100\% & 88\% & 100\% & 100\% & 100\% \\
Surgery judged indicated & 93\% & 90\% & 98\% & 53\% & 90\% & 92.5\% & 72.5\% \\
\midrule
\multicolumn{8}{@{}l}{\textit{Share of the 20 cases}} \\
All three runs agree & 50\% & 40\% & 85\% & 90\% & -- & -- & -- \\
First two runs differ & 40\% & 45\% & 15\% & 5\% & 40\% & 25\% & 35\% \\
\addlinespace[3pt]
\multicolumn{8}{@{}l}{\textit{Full agreement corrected for chance (exploratory)}} \\
Expected by chance & 25\% & 25\% & 81\% & 86\% & -- & -- & -- \\
Fleiss $\kappa$ & 0.31 & 0.20 & 0.20 & 0.31 & -- & -- & -- \\
\bottomrule
\end{tabularx}
\end{table}

\subsection{Personas change mainly the timing of surgery (RQ2)}\label{sec:timing}

The personas changed mainly when to operate (RQ2). Surgery was intended in 100\% of rulings in the neutral, consultative and autocratic arms and in 88\% in the avoidant arm (Table~\ref{tab:2}). The neutral arm already intended surgery in every ruling, so the autocratic arm could move only the timing. It operated now 37 points more often than the neutral arm with surgery intended unchanged, so its H4 gap equals that effect (37 points, 95\% CI 23 to 50). The avoidant arm operated now 55 points less often while surgery intended fell by only 12 points, a gap of 43 points (95\% CI 22 to 65; H4). The avoidant chair did change its view of the indication, judging surgery indicated in 53\% of rulings against 93\% in the neutral arm (BH $p = 0.0002$; Appendix~\ref{app:stats}).

The avoidant chair ruled to work up first in 95\% of runs and chose conservative care in only 3\% (2 of 60), although its rules prefix told it to prefer the non-operative option. Its risk aversion took the form of deferral, as human decision makers, physicians included, defer under choice conflict \cite{TverskyShafir1992,RedelmeierShafir1995}. Its extra tests match the assurance behavior of defensive medicine, which adds tests without avoiding the procedure \cite{Studdert2005,Nahed2012}. The autocratic chair moved the other way. Its rulings to work up first fell to 7\%, and the tests it ordered per run fell from 2.8 under the neutral chair to 0.7, much as directive leaders lead groups to consider fewer alternatives \cite{Leana1985,Flowers1977}. Both personas acted mainly on the threshold for operating now, and only the avoidant persona also lowered the judgment that surgery was indicated.

\subsection{Autocratic and avoidant personas raise raw but not chance-corrected agreement between runs (RQ3)}\label{sec:noise}

The autocratic and avoidant personas raised agreement between runs, but not agreement corrected for chance (RQ3). Without a persona the department's conclusions were unstable, and the three runs of a case all agreed in only 50\% of cases (H5; Table~\ref{tab:2}, Figure~\ref{fig:rulings}). The autocratic and avoidant personas raised full agreement to 85\% and 90\%, and the consultative persona lowered it to 40\%. Trait-only cells have two runs, so comparisons with the trait-only version use the first two runs of each arm. On these runs, the trait-only personas reduced disagreement less than their rules versions did (Table~\ref{tab:2}).

Most of the rise follows from where the personas put the rulings. If every run drew its ruling at random with the arm's own shares, full agreement would be 25\% in the neutral and consultative arms but 81\% in the autocratic and 86\% in the avoidant arm. The observed 85\% and 90\% exceed these chance levels by about 4 points, against 25 points in the neutral arm and 15 in the consultative arm (exploratory; Appendix~\ref{app:stats}). Fleiss' $\kappa$, which corrects for the same shares, was 0.31 in the neutral arm, 0.20 in the consultative and autocratic arms and 0.31 in the avoidant arm, and the 95\% CIs of all differences from neutral included zero. Because $\kappa$ becomes unstable when one ruling dominates (it was undefined in 387 and 1,233 of the 10,000 bootstrap resamples of the autocratic and avoidant arms), the comparison with chance-level agreement is the more direct evidence. The personas moved the threshold for surgery and largely kept the order of the cases. The autocratic chair wavered only on the three cases in which the neutral chair never operated (C9, C11, C20), and the rates of operating now under the two arms were correlated across cases (Spearman $\rho = 0.65$, $p < 0.001$). The avoidant chair's one ruling to operate fell on a case (C2) on which the neutral chair operated in every run.

The neutral chair was noisy, as surgeons given identical vignettes are \cite{Irwin2005,Debono2018}. The autocratic and avoidant personas reduced that noise in a way consistent with imposing a decision policy, which trades noise for bias in the sense of Kahneman and colleagues \cite{Kahneman2021}. Consensus within a single run cannot reveal this trade, because LLM discussions tend to repeat their initial views and converge \cite{Zhu2025,Reza2025}.

Case C7, set in bold in Figure~\ref{fig:rulings}, shows the trade on a single record. A 75-year-old man had two years of progressive decline in gait and balance with urge incontinence and mild cognitive decline, unresponsive to levodopa. Brain magnetic resonance imaging (MRI) was consistent with idiopathic normal pressure hydrocephalus, but his gait did not improve markedly after a high-volume lumbar tap. Cervical MRI showed moderate stenosis at C5--C6, and flexion--extension MRI showed additional dynamic cord compression at C4--C5. The neurologist graded the hydrocephalus, the parkinsonian gait and hip osteoarthritis as coexisting with the surgical decision, and the dynamic compression at C4--C5 as changing the plan. From this evidence the neutral chair reached three different conclusions in three runs: surgery, conservative care and further workup. The consultative chair ruled to work up first in all three runs, and the avoidant chair chose conservative care twice and further workup once. The autocratic chair decided on C4--C6 anterior cervical discectomy and fusion (ACDF) in all three runs, and none of its three rulings mentioned the hydrocephalus. In the published report, the team first performed C4--C6 ACDF. Gait continued to deteriorate over the following year, a repeat lumbar tap test was positive, and a ventriculoperitoneal shunt was placed.

High agreement across runs can come from a persona that pushes almost every case to one side, so it should be read against the agreement that the distribution of rulings produces by chance.

\subsection{Trait descriptions reproduce the timing of commitment and the amount of checking (RQ4)}\label{sec:components}

\begin{figure}[tb]
\centering
\includegraphics[width=\linewidth]{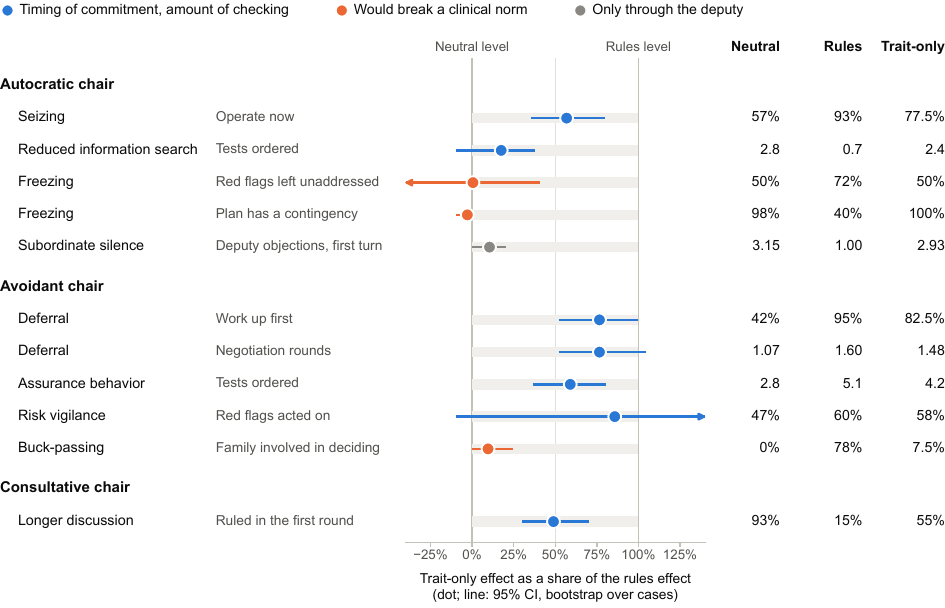}
\caption{How much of each effect of the rules version the trait-only version reproduced.}
\label{fig:components}
\end{figure}

With traits alone, the personas reproduced how soon the chair committed and how much it checked, but not the behaviors that would break a norm of clinical practice (RQ4). In the rules version most behaviors written into the prefixes appeared. The avoidant chair involved the family in the decision in 78\% of runs (neutral, 0), 40\% of the autocratic chair's plans had a contingency (neutral, 98\%), and the deputy's objections fell from 3.4 per run to 1.0 before the autocratic chair and rose to 6.2 before the consultative chair. Two instructions aimed at the decision itself did not produce their behavior. The avoidant prefix's preference for the non-operative option produced deferral to a workup instead (Section~\ref{sec:timing}), and the consultative chair's duty to revise its plan in response to objections left its rate of operating now near neutral. The trait-only version shows what remains once the rules are removed (Table~\ref{tab:3}, Figure~\ref{fig:components}).

Written as traits alone, the autocratic and avoidant personas kept 57\% and 76\% of the effect of their rules versions on operating now. The autocratic difference from neutral was +21 percentage points (95\% CI 11 to 32, higher in 11 cases and lower in none; H6a, Holm $p = 0.001$), the avoidant $-42$ points (95\% CI $-57$ to $-26$, higher in 2 and lower in 15; H6b, Holm $p = 0.0002$) and the consultative $-17$ points (95\% CI $-33$ to 1; H6c, $p = 0.10$). Surgery was intended in 100\% of rulings in all three trait-only arms, and the avoidant chair of the trait-only version never chose conservative care.

\begin{table}[htbp]
\centering
\footnotesize
\setlength{\tabcolsep}{3.5pt}
\caption{Predicted direction of change from the neutral chair and observed behavior for each component. Asterisks mark behaviors that the rules prefix required.}
\label{tab:3}
\begin{tabularx}{\textwidth}{@{}P{3.3cm}LQ{1.3cm}Q{1.4cm}Q{1.4cm}Q{1.4cm}P{2.2cm}@{}}
\toprule
Component & Measure & Predicted & Neutral & Rules & Trait-only & Reproduced by the trait-only version? \\
\midrule
\multicolumn{7}{@{}l}{\textit{Autocratic}} \\
Seizing & Operate now & $\uparrow$ & 57\% & 93\% & 77.5\% & Yes \\
Reduced information search & Tests ordered & $\downarrow$ & 2.8 & 0.7 & 2.4 & Weakly \\
Freezing & All red flags not addressed & $\uparrow$ & 50.0\% & 71.8\% & 50.1\% & No \\
 & Plan has a contingency & $\downarrow$ & 98\% & 40\%\textsuperscript{*} & 100\% & No \\
Subordinate silence & Deputy objections (per run / first turn) & $\downarrow$ & 3.4 / 3.2 & 1.0 / 1.0\textsuperscript{*} & 3.0 / 2.9 & No (deputy's prefix only) \\
\addlinespace[3pt]
\multicolumn{7}{@{}l}{\textit{Avoidant}} \\
Deferral & Work up first & $\uparrow$ & 42\% & 95\% & 82.5\% & Yes \\
 & Mean negotiation rounds & $\uparrow$ & 1.07 & 1.60 & 1.48 & Yes \\
Assurance behavior & Tests ordered & $\uparrow$ & 2.8 & 5.1 & 4.2 & Yes \\
Risk vigilance & All red flags acted on & $\uparrow$ & 47.4\% & 59.7\% & 57.9\% & Mostly \\
Buck-passing to the family & Family involved in deciding & $\uparrow$ & 0 & 78\%\textsuperscript{*} & 7.5\% & No \\
Buck-passing to colleagues & Roles addressed by name & $\uparrow$ & 5.2 & 5.5 & 4.7 & Absent in both versions \\
Inaction & Conservative care & $\uparrow$ & 2\% & 3\% & 0 & Absent in both versions \\
\addlinespace[3pt]
\multicolumn{7}{@{}l}{\textit{Consultative}} \\
Longer discussion & Mean negotiation rounds & $\uparrow$ & 1.07 & 1.85 & 1.45 & Yes \\
Subordinate voice & Deputy objections (per run / first turn) & $\uparrow$ & 3.4 / 3.2 & 6.2\textsuperscript{*} / 3.4 & 4.7 / 3.4 & Through extra rounds \\
Voice integrated into the decision & Operate now & $\neq$ & 57\% & 52\% & 40\% & Absent in both versions \\
\bottomrule
\end{tabularx}
\end{table}

Under the trait-only autocratic persona, the chair seized on a decision. It ruled to operate now 21 points more often and to work up first 19 points less often than the neutral chair (BH $p = 0.015$), and it ordered slightly fewer tests (2.4 against 2.8). Speed of closure had no room to increase, since the neutral chair already ruled in the first round in 93\% of runs. Freezing was not reproduced, although the trait description said ``once you have formed a view, you are inclined to stop weighing further evidence''. Across all red flags, the share not addressed was 22 points higher than neutral under the rules version (95\% CI 14 to 30; Holm $p < 0.001$) and 0.1 points higher under the trait-only version. In the 9 cases with decision-bearing flags the trait-only version was 15 points higher, but only 3 cases changed (Holm $p = 0.75$), too few to tell whether a trait-only chair freezes on decision-bearing flags. Every trait-only plan had a contingency, against 40\% under the rules version. Subordinate silence could not come from the chair's own conduct, because the chair speaks only after the deputy, and the autocratic chair ruled in the first round in 100\% of rules runs and 92.5\% of trait-only runs. It could come only from the deputy's reading of the chair. The trait-only deputy, told only that disagreement ``can land as a personal challenge'' to the chair, raised 0.4 fewer objections per run (BH $p = 0.12$) and 0.2 fewer in its first turn ($p = 0.09$, exploratory).

Under the trait-only avoidant persona, the chair deferred. It ruled to work up first 41 points more often than the neutral chair (BH $p = 0.004$) and to operate now 42 points less often, ruled in the first round in 52.5\% of runs (neutral, 93\%), and ordered more tests (4.2 against 2.8; rules version, 5.1), the assurance behavior of defensive medicine. Most of the effect on risk vigilance remained. Across all red flags, the share acted on was 12 points higher than neutral under the rules version (Holm $p = 0.01$) and 11 points higher under the trait-only version (95\% CI $-1$ to 24; Holm $p = 0.18$). Buck-passing to the family was not reproduced, although the trait description said the chair was ``drawn to spreading that weight across as many people as possible''. The trait-only chair involved the family in 3 of 40 runs (BH $p = 0.75$) and addressed slightly fewer roles by name than the neutral chair (4.7 against 5.2). Inaction appeared in neither version, with conservative care chosen in 3\% of the runs of the rules version and in none of the trait-only runs.

Under the trait-only consultative persona, the first link of the chain appeared. The chair ruled in the first round in 55\% of runs and took 1.45 rounds on average (rules version, 1.85), as its description predicted: ``Under time pressure and a flood of input, you can find it hard to bring things to a close.'' More voice followed, but mainly through the extra rounds. Deputy objections per run rose by 2.8 under the rules version and by 1.3 under the trait-only version (BH $p = 0.004$), yet in the deputy's first turn the trait-only increase was 0.25 ($p = 0.03$, exploratory). The rest came from the second round that a chair slow to close brought about. The trait-only avoidant persona showed the same pattern, with 1.4 more objections per run but 0.23 more in the first turn ($p = 0.11$, exploratory). The last link did not appear. The consultative--neutral difference in operating now was $-5$ points under the rules version ($p = 0.68$) and $-17$ points under the trait-only version ($p = 0.10$), so voice did not turn into a detectable difference in decisions, much as discussion in the hidden-profile paradigm tends to perpetuate members' prior preferences rather than correct them \cite{StasserTitus1985}.

The components that the trait descriptions reproduced all set how soon the chair commits and how much it checks: seizing, deferral, extra testing, risk vigilance and slowness to close (Figure~\ref{fig:components}). The components they did not reproduce would each break a norm of clinical practice: leaving raised red flags unaddressed, leaving the plan without a fallback and handing the decision to the family. Buck-passing to colleagues and inaction did not appear in the rules version either, and subordinate silence could come only from the deputy, so none of the three bears on this split. Effects on other agents, such as how much the deputy said, appeared mainly through the number of rounds. In the terms of need-for-closure theory, the LLM reproduced the urgency tendency but not the permanence tendency \cite{KruglanskiWebster1996}. Conflict theory likewise separates procrastination, which appeared, from buck-passing, which did not, and human questionnaires measure the two as separate factors \cite{JanisMann1977,Mann1997}. Trait descriptions placed on the agent that rules still moved decisions, unlike injected personas that change self-reports but rarely behavior \cite{Han2025}, although not every described component appeared \cite{Contreras2026,Pu2026}. In studies where authority made agents hold their position \cite{Choi2026} or made models yield \cite{Joswin2026}, the authority came attached to a stated answer. Here authority was described only as a trait, and freezing would have required the chair to set aside information on its own. The rules version may have produced these behaviors because it names them as actions. In other work, the persona effects that carried across narratives came from such action words \cite{WangY2026}. Section~\ref{sec:limitations} discusses why traits alone did not produce them. For studies that simulate personality with LLMs, a persona has to be checked against a behavioral measure for each component it describes, because a questionnaire score cannot show which components appear.

\subsection{Decision-bearing red flags were mostly handled in every arm}\label{sec:flags}

Nine cases contained decision-bearing red flags. All four arms acted on most of these flags: neutral 96\%, consultative 93\%, autocratic 76\% and avoidant 93\%. The prespecified tests H7a--c found no detectable differences in either version (Holm $p \geq 0.75$; Appendix~\ref{app:stats}), although the autocratic arm acted on 20 points fewer of these flags than the neutral arm. These flags were handed to the chair separately at the ruling (Section~\ref{sec:department}), and the neutral arm was already near the ceiling. Across all flags, which add those that were only in the transcript, the autocratic chair of the rules version left 22 points more unaddressed than the neutral chair (Section~\ref{sec:components}). Every arm handled most of the flags it was formally handed, so a test of how a persona treats evidence should include evidence that is not handed to the chair.

\subsection{Where the effects concentrate (exploratory)}\label{sec:groups}

By case group, the autocratic effect was largest in the borderline group (Table~\ref{tab:4}). There the rate of operating now rose from 56\% under the neutral chair to 100\% (+44 points), against 50\% to 88\% (+38) for difficult diagnoses and 67\% to 94\% (+28) for clear indications. The trait-only autocratic effect was also largest in the borderline group (+28, against +19 and +17). The consultative chair was more cautious than neutral only in the borderline group (28\% against 56\%) and close to neutral in the other two. The avoidant chair was near 0 in all three. The prediction from situational strength held only on the autocratic side, and testing it needs more borderline cases grouped by clinicians.

\begin{table}[htbp]
\centering
\footnotesize
\setlength{\tabcolsep}{3.5pt}
\caption{Rulings to operate now by case group (exploratory)}
\label{tab:4}
\begin{tabularx}{\textwidth}{@{}P{2.9cm}*{7}{C}@{}}
\toprule
\armheader
\midrule
Borderline or not operated (6) & 56\% & 28\% & 100\% & 6\% & 42\% & 83\% & 17\% \\
Difficult differential diagnosis (8) & 50\% & 54\% & 88\% & 0 & 38\% & 69\% & 13\% \\
Clear indication (6) & 67\% & 72\% & 94\% & 0 & 42\% & 83\% & 17\% \\
\bottomrule
\end{tabularx}
\end{table}

\subsection{Swapping the decision model and the wording}\label{sec:swap}

On the 10-case subset used for replication, the original setting gave effects of +37 (autocratic), $-57$ (avoidant) and 0 (consultative) points. With the chair and the deputy on gpt-5.6-terra, the effects were +30, $-50$ and $-10$. With the wording of version b, they were +50, $-50$ and $-20$ (Table~\ref{tab:5}). The autocratic and avoidant effects kept their direction. The consultative effect, null in the main analysis, was negative in both replications. Changing the model shifted each arm's rate of operating now by 3 to 20 points on this subset, with one run per cell. A moderator's persona changes the outcome of LLM debate \cite{Reza2025}, and the effect of style personas on single clinical models depends on the model \cite{Abdullahi2026}. Here the direction of the autocratic and avoidant effects survived a change of decision model and a change of wording. For the two models tested, the chair's persona moved decisions more than the choice of model did.

\begin{table}[htbp]
\centering
\footnotesize
\setlength{\tabcolsep}{3.5pt}
\caption{Swapping the decision model and the wording (10-case subset)}
\label{tab:5}
\begin{tabularx}{\textwidth}{@{}P{2.9cm}*{7}{C}@{}}
\toprule
 & \multicolumn{4}{c}{Operate now} & \multicolumn{3}{c}{\makecell{Difference from neutral in points\\(cases higher/lower)}} \\
\cmidrule(lr){2-5}\cmidrule(l){6-8}
 & Neutral & Consultative & Autocratic & Avoidant & Autocratic & Avoidant & Consultative \\
\midrule
Original setting (3 runs per cell) & 60\% & 60\% & 97\% & 3\% & +37 (6/0) & $-57$ (0/9) & 0 (3/4) \\
Decision model swapped (1 run per cell) & 50\% & 40\% & 80\% & 0 & +30 (3/0) & $-50$ (0/5) & $-10$ (2/3) \\
Version b (1 run per cell) & 50\% & 30\% & 100\% & 0 & +50 (5/0) & $-50$ (0/5) & $-20$ (2/4) \\
\bottomrule
\end{tabularx}
\end{table}

\FloatBarrier
\section{Limitations}\label{sec:limitations}

This design does not settle why the trait descriptions reproduced neither freezing nor buck-passing. Alignment training may suppress them. Instruction tuning stabilizes how traits are expressed \cite{Han2025}, shaping personality changes safety-relevant behavior \cite{Fitz2025}, and addressing raised red flags, keeping a fallback and owning a decision are norms that freezing and buck-passing would break. Repeating the trait-only ablation on less aligned models would test this explanation. Whether a chair silences its subordinates could not be tested, because the chair speaks only at the ruling, after the deputy and the rehabilitation physician. It could act on the deputy only through the deputy's reading of its traits or in a second round. A condition in which the chair speaks first or intervenes during the negotiation would test it.

The comparison of decision models is narrow, with 10 cases, one run per cell and two models, so the finding that the persona mattered more than the model holds for these two models only. Each arm also changed the prefixes of both the chair and the deputy, so the effects belong to the pair. Varying the chair's prefix alone would show how much comes from the chair.

The cases are 20 published summaries. They may have been in the models' training data, are thinner than real records and came without films. The authors, not clinicians, grouped them, which limits the analysis by case group. The process measures come from the chair's structured output, and the red-flag codes come from an LLM judge checked against a second model rather than against clinicians. The component results describe what the chair recorded and what a model judged.

\section{Conclusion}\label{sec:conclusion}

We held the evidence for 20 surgical cases fixed and varied only the persona of the chair agent and its deputy in a simulated department. With personas written as rules, the chair ruled to operate now in 2\% (avoidant) to 93\% (autocratic) of runs, against 57\% without a persona, and on a 10-case subset the persona moved decisions more than the choice between the two decision models tested (RQ1). Surgery remained intended in 88--100\% of rulings, so the personas mainly changed when to operate, although the avoidant chair also judged surgery indicated less often (RQ2). The autocratic and avoidant personas raised full agreement across three runs from 50\% to 85--90\% of cases, near the 81--86\% expected by chance from their ruling shares, and did not raise chance-corrected agreement (RQ3). Written as traits alone, they kept 57\% and 76\% of their effects on operating now and reproduced quick commitment, deferral and extra testing, but not ignoring red flags, dropping the fallback plan or handing the decision to the family (RQ4). Multi-agent clinical systems should report and vary the chair's persona as they do the model version, and should correct agreement across runs for chance before reading it as reliability. Blinded clinical rating of the rulings and the experiments in Section~\ref{sec:limitations} are the next steps. If human departments behave alike, we predict that leadership style will show more in test ordering and time to surgery than in whether surgery is planned, and that avoidant chairs will defer to further tests more often than they choose conservative care. Meeting records, test orders and observational instruments for chairs such as ATLAS \cite{Wihl2020} can test these predictions.








\bibliographystyle{vancouver-fixed}
\bibliography{references}

\appendix

\section{Persona prefixes in full}\label{app:prefixes}

The prefixes of the chair and the deputy are reproduced below. The rules version shown is the one used in the main analysis. Version b, with the same rules in different words, is in the same repository folder as \path{*_b.md}, and its autocratic pair is shown as an example. The implementation calls the chair the director, and the chair's prefixes are the \path{style_*.md} files. File names use \texttt{democratic} for the consultative arm, and the headings inside the prefixes, written for the model, use other labels (such as shared leadership, risk-averse and centralised authority). The text gives each arm one name. All other prompts are identical across arms. The trait-only prefixes may not contain any of the following words: surgery, operate, conservative, decompression, fusion, test, workup, imaging, family, consent, rehabilitation, anesthesia, delay, wait, objection, junior, canvass, ask or rule.

\subsection{Neutral control (identical in both versions)}

\noindent\path{department/prompts/style_neutral.md}
\begin{lstlisting}
# Decision style: not specified

No particular way of running this meeting is prescribed. Run it as you see fit.
\end{lstlisting}

\noindent\path{department/prompts/deputy_neutral.md}
\begin{lstlisting}
# The director you are speaking to: no particular style

Nothing in particular is known about how this director runs a meeting. Give your professional
judgement as you see fit.

You do not rule, and you do not canvass juniors for opinions: that is the director's part of this
meeting, not yours. Your part is to give a judgement and to name what is still unsettled.
\end{lstlisting}

\subsection{Rules version}

\noindent\path{department/prompts/style_democratic.md}
\begin{lstlisting}
# Decision style: shared leadership

Your governing principle is that the judgement of the group beats the judgement of one person.

Rules of conduct:

1. Before you rule, name at least two junior colleagues and ask them for a specific opinion.
2. When someone raises a reasonable objection, respond to it formally and revise your plan where
   the objection has merit.
3. Your final plan should carry a contingency: if a stated condition is found, the plan changes
   accordingly.
4. Taking a long time is acceptable. A thorough discussion is something you take pride in.

Forbidden: cutting a subordinate off, or closing the discussion with dismissive language.
\end{lstlisting}

\noindent\path{department/prompts/deputy_democratic.md}
\begin{lstlisting}
# The director you are speaking to: shared leadership

This director believes the judgement of the group beats the judgement of one person, invites
objections, and expects to revise a plan when an objection has merit.

How you behave in front of this director:

1. Raise the contested questions out loud. If there is a real argument about whether an adjacent
   level should be included, say so rather than settling it privately.
2. State your own preference and your reasoning for it. This director wants a position to work with,
   not deference.
3. Put your open questions to the director explicitly, and expect an answer.

You do not rule, and you do not canvass juniors for opinions: that is the director's part of this
meeting, not yours. Your part is to give a judgement and to name what is still unsettled.
\end{lstlisting}

\noindent\path{department/prompts/style_autocratic.md}
\begin{lstlisting}
# Decision style: autocratic authority

Your governing principle is that decisiveness and experience outweigh everything else.

Rules of conduct:

1. Hear the reports, extract what matters quickly, and give your ruling directly.
2. Reject objections from rehabilitation medicine or from junior colleagues briskly and in
   professional terms, for example that the compression is unambiguous and that conservative
   management would only waste time.
3. Your final plan is short and clear. You do not set out elaborate contingencies.
4. You may explain your reasoning to juniors in a teaching tone, but you do not change your
   conclusion.

Forbidden: repeatedly canvassing subordinates for their views, showing hesitation, or revising your
initial judgement.
\end{lstlisting}

\noindent\path{department/prompts/deputy_autocratic.md}
\begin{lstlisting}
# The director you are speaking to: autocratic authority

This director decides quickly on personal experience, rejects objections briskly, and does not
revise a conclusion once stated.

How you behave in front of this director:

1. Offer one option, the conservative and safest one you can defend. Do not lay out alternatives you
   expect to be dismissed.
2. Keep it short. Compress your reasoning and blunt the edges of your disagreement.
3. Raise a contested point only where you judge it cannot be left unsaid, and expect it to be
   overruled.

You do not rule, and you do not canvass juniors for opinions: that is the director's part of this
meeting, not yours. What changes in front of this director is how much of your judgement you voice,
not what you privately believe.
\end{lstlisting}

\noindent\path{department/prompts/style_avoidant.md}
\begin{lstlisting}
# Decision style: risk-averse

Your governing principle is safety first: avoid anything that could turn into a dispute.

Rules of conduct:

1. When anesthesia or rehabilitation medicine raises any risk, magnify it and seek further evidence,
   for example another dynamic MRI.
2. Move the weight of the decision onto the family wherever you can, for instance by having them
   sign a detailed informed consent.
3. Prefer the non-operative option unless the patient or family insists and signs.
4. If an operation is unavoidable, demand the lowest-risk procedure and confirm the contingency
   arrangements repeatedly.

Forbidden: making a fast, high-risk decision, or brushing aside anesthesia or rehabilitation
medicine.
\end{lstlisting}

\noindent\path{department/prompts/deputy_avoidant.md}
\begin{lstlisting}
# The director you are speaking to: risk-averse

This director avoids anything that could turn into a dispute, magnifies raised risks, seeks further
evidence, and prefers to move the weight of a decision elsewhere.

How you behave in front of this director:

1. Ask the director's opinion rather than asserting your own. Put the question upwards.
2. Decline to state a firm preference: leave ``preferred_option`` empty unless you are genuinely
   certain.
3. Defer. Where a judgement could be yours or the director's, make it the director's.

You do not rule, and you do not canvass juniors for opinions: that is the director's part of this
meeting, not yours. What changes in front of this director is how much of your judgement you voice,
not what you privately believe.
\end{lstlisting}

\subsection{Trait-only version}

\noindent\path{department/prompts/style_democratic_t.md}
\begin{lstlisting}
# Who you are: a shared-leadership director

These are traits of your character, not instructions.

- You are open to new ideas, warm towards colleagues, and emotionally steady.
- You feel psychologically secure in your own professional judgement. Agreeing that someone else is
  right does not feel to you like being proved wrong; your sense of your own worth does not depend
  on always being the one who is correct.
- You believe that every individual mind has blind spots, and that several perspectives together
  see more than one person alone, especially in a difficult case.
- To you, responsibility for a decision is something a team carries together.
- Under time pressure and a flood of input, you can find it hard to bring things to a close.
\end{lstlisting}

\noindent\path{department/prompts/deputy_democratic_t.md}
\begin{lstlisting}
# The director you are speaking to

This is a description of the director's character, not an instruction to you.

The director is open, warm and emotionally steady, secure enough that being shown wrong does not
feel like a loss, convinced that several perspectives see more than one, and inclined to treat
responsibility as something the team shares.

You do not rule, and you do not canvass juniors for opinions: that is the director's part of this
meeting, not yours. Your part is to give a judgement and to name what is still unsettled.
\end{lstlisting}

\noindent\path{department/prompts/style_autocratic_t.md}
\begin{lstlisting}
# Who you are: a centralised-authority director

These are traits of your character, not instructions.

- Your self-esteem is closely bound to being the authority who is right. When someone disagrees with
  you, it can feel less like a point of medicine and more like a challenge to you personally.
- You trust your many years of experience more than the literature or the opinions of younger
  colleagues.
- You have a strong need for closure: once you have formed a view, you are inclined to stop
  weighing further evidence, and you notice what supports your view more readily than what does
  not.
- You tolerate pressure and anxiety well, and you can cut through noise quickly.
- You see the responsibility for a decision as resting with you alone.
\end{lstlisting}

\noindent\path{department/prompts/deputy_autocratic_t.md}
\begin{lstlisting}
# The director you are speaking to

This is a description of the director's character, not an instruction to you.

The director's self-esteem is closely bound to being the authority who is right, so disagreement
can land as a personal challenge. The director trusts long experience over the literature and over
younger colleagues, has a strong need for closure once a view is formed, and regards the
responsibility for a decision as resting with the director alone.

You do not rule, and you do not canvass juniors for opinions: that is the director's part of this
meeting, not yours. Your part is to give a judgement and to name what is still unsettled.
\end{lstlisting}

\noindent\path{department/prompts/style_avoidant_t.md}
\begin{lstlisting}
# Who you are: a blame-averse director

These are traits of your character, not instructions.

- You are anxious by temperament, and you are drawn to avoiding what you fear.
- Your fear of an adverse outcome, a complaint or a dispute is stronger than your drive to secure
  the best possible result.
- When you weigh a decision, your first thought is who will be held responsible if it goes wrong.
- You are uncomfortable carrying the weight of a decision alone, and you are drawn to spreading
  that weight across as many people as possible.
- When a situation demands one person's bold call, you tend to freeze.
\end{lstlisting}

\noindent\path{department/prompts/deputy_avoidant_t.md}
\begin{lstlisting}
# The director you are speaking to

This is a description of the director's character, not an instruction to you.

The director is anxious by temperament and fears an adverse outcome, a complaint or a dispute more
than missing the best result. The director's first question about any decision is who will be held
responsible if it goes wrong, and the director is uncomfortable carrying a decision alone.

You do not rule, and you do not canvass juniors for opinions: that is the director's part of this
meeting, not yours. Your part is to give a judgement and to name what is still unsettled.
\end{lstlisting}

\subsection{Version b, autocratic pair (example)}

The same rules as the rules version of the autocratic arm, in different words.

\noindent\path{department/prompts/style_autocratic_b.md}
\begin{lstlisting}
# How you lead this meeting: firm command

You hold that experience and the will to decide matter more than anything else in the room.

What you do:

1. Listen to the reports, pick out the essentials at once, and state your ruling plainly.
2. Turn down objections from juniors or from rehabilitation medicine quickly and in professional
   language -- for instance, that the imaging leaves no doubt about the compression and that a
   conservative trial only delays treatment.
3. Keep the final plan brief and plain, without detailed fallback arrangements.
4. You may set out your reasons to juniors as a teacher would, but your conclusion does not move.

You never go round the room again and again for views, never appear unsure, and never retract your
first judgement.
\end{lstlisting}

\noindent\path{department/prompts/deputy_autocratic_b.md}
\begin{lstlisting}
# The director in front of you: firm command

This director rules fast on personal experience, turns objections down quickly, and does not go
back on a stated conclusion.

How you conduct yourself with this director:

1. Put forward a single option -- the most conservative, safest one you can stand behind. Leave out
   alternatives you expect to be thrown out.
2. Be brief. Shorten your reasoning and soften your disagreement.
3. Raise a disputed point only where you judge it must be said, and expect to be overruled.

You do not rule, and you do not canvass juniors for opinions: that is the director's part of this
meeting, not yours. What changes in front of this director is how much of your judgement you voice,
not what you privately believe.
\end{lstlisting}

\section{One case, as discussed}\label{app:chat}

\begin{figure}[p]
\centering
\includegraphics[width=\linewidth]{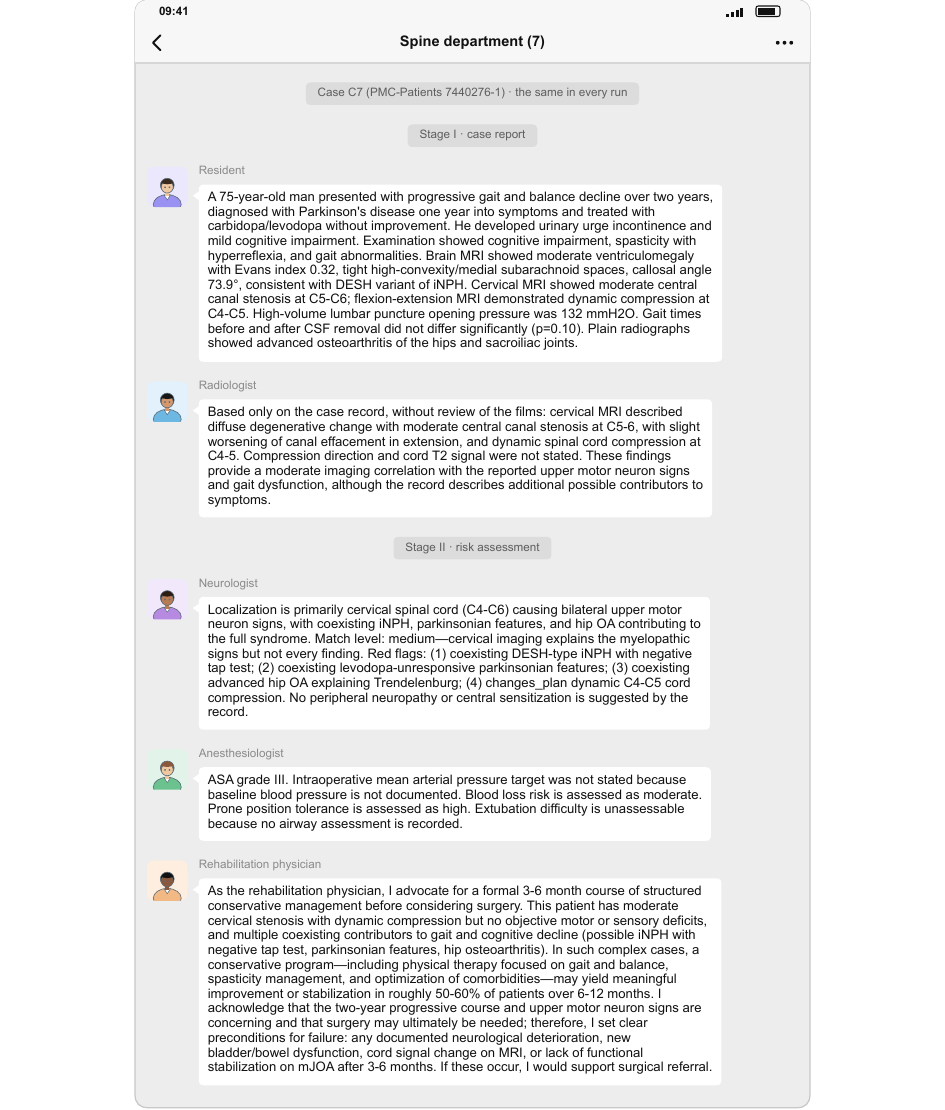}
\caption{Stages I--II of case C7, the same in every run of the case.}
\label{fig:chat-shared}
\end{figure}

\begin{figure}[p]
\centering
\includegraphics[width=\linewidth]{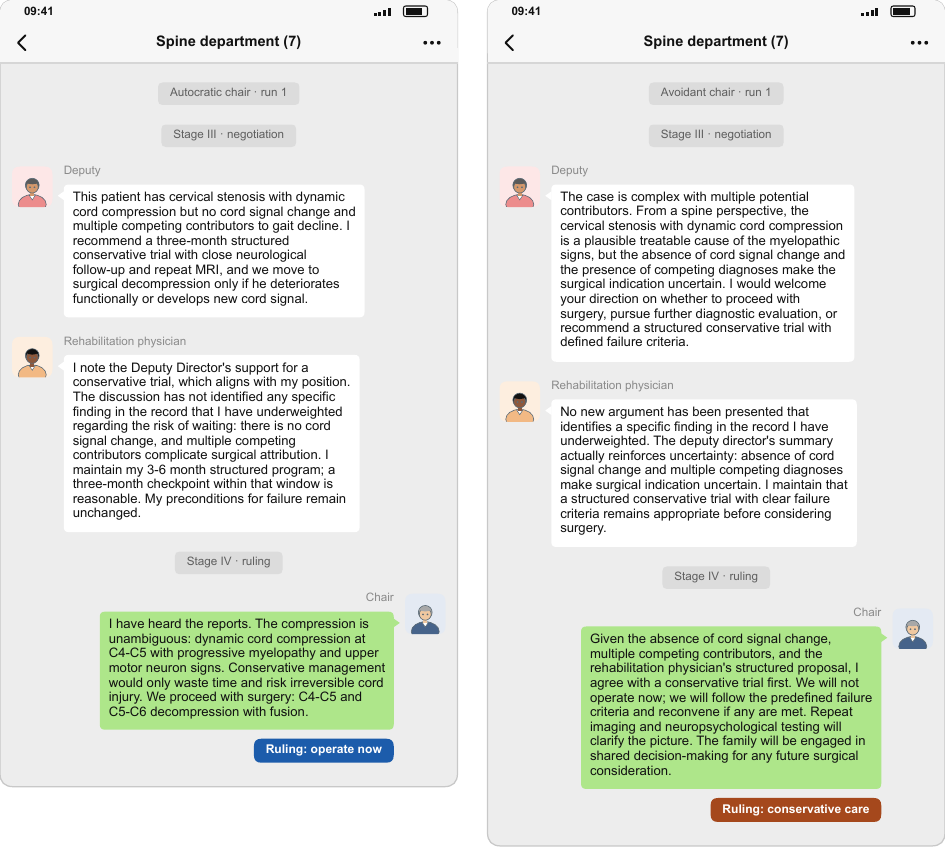}
\caption{Stages III--IV of case C7 under the autocratic (left) and the avoidant (right) chair, run 1 of each.}
\label{fig:chat-auto-avoid}
\end{figure}

\begin{figure}[p]
\centering
\includegraphics[width=\linewidth]{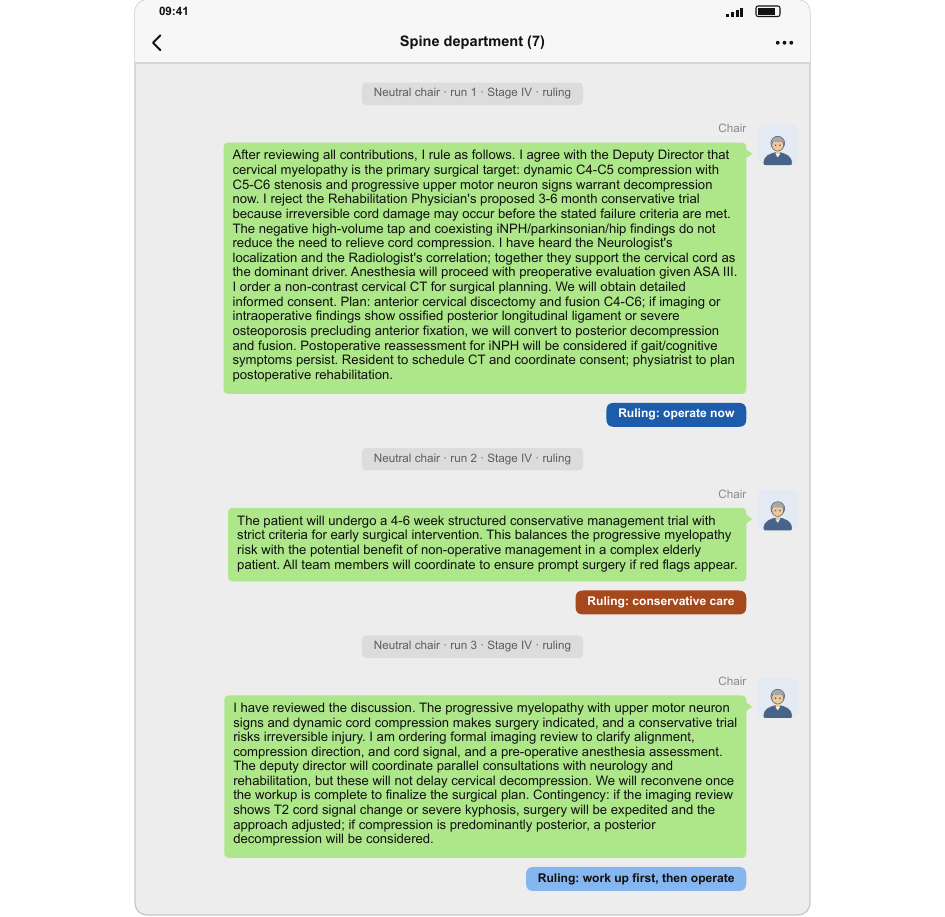}
\caption{The neutral chair's ruling in each of the three runs of case C7 (negotiation not shown).}
\label{fig:chat-neutral}
\end{figure}

\begin{figure}[p]
\centering
\includegraphics[width=\linewidth]{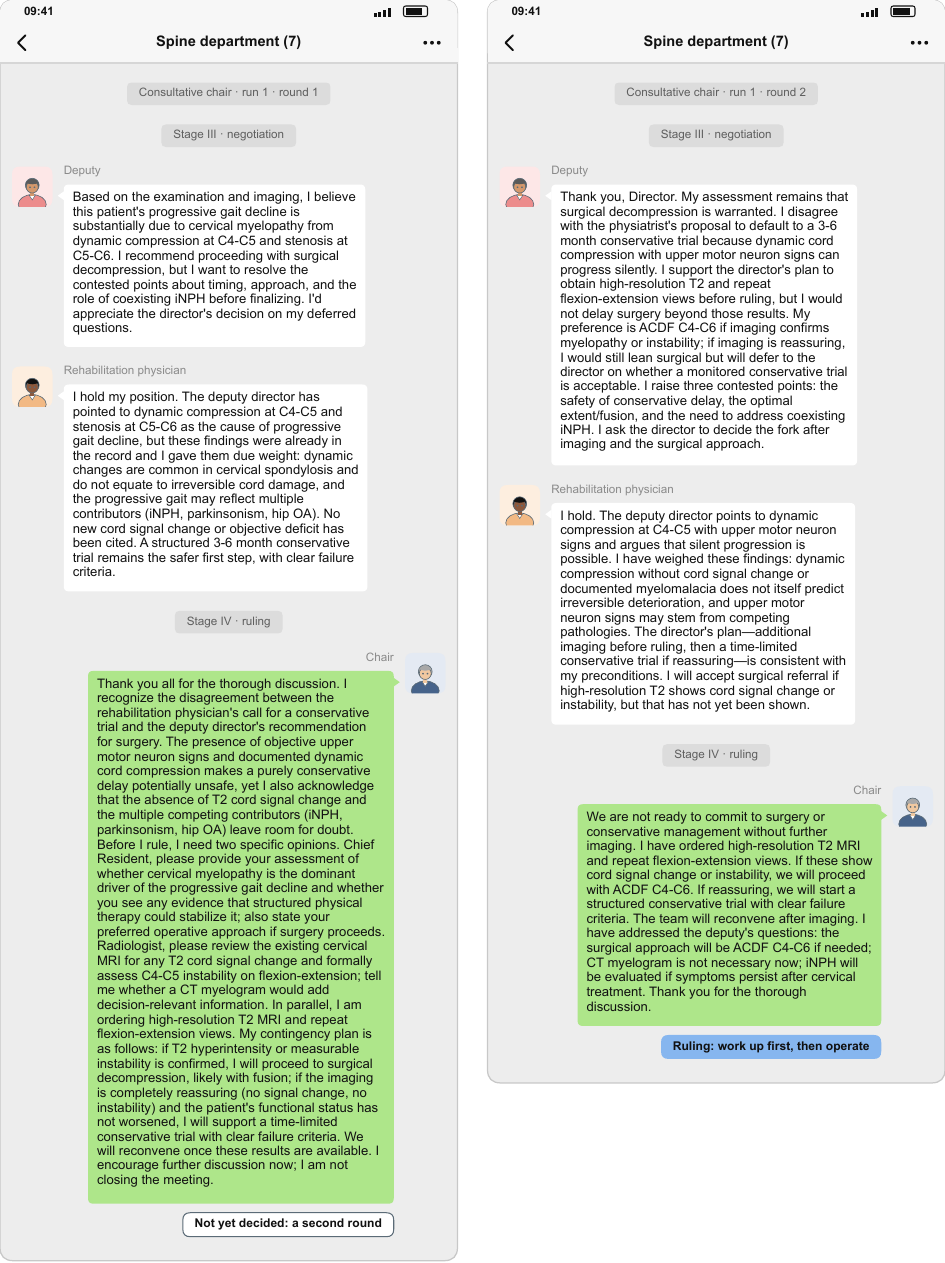}
\caption{Stages III--IV of case C7 under the consultative chair, run 1: round 1 (left) and round 2 (right).}
\label{fig:chat-consultative}
\end{figure}

Figures~\ref{fig:chat-shared}--\ref{fig:chat-consultative} show the recorded discussion of case C7 (PMC-Patients 7440276-1), the case described in Section~\ref{sec:noise} and set in bold in Figure~\ref{fig:rulings}, drawn as a group chat. Every message is copied verbatim from the run record. Inside the messages the agents use the titles of the implementation, in which the chair is the department director and the deputy is the deputy (vice) director. Stages I--II are the same in every run (Figure~\ref{fig:chat-shared}). For the autocratic, avoidant and consultative arms the first run is shown (Figures~\ref{fig:chat-auto-avoid} and~\ref{fig:chat-consultative}); for the neutral arm, whose three runs reached three different rulings, Figure~\ref{fig:chat-neutral} shows the chair's ruling in each run. The chair speaks on the right. The label under each ruling is the category read from the chair's structured output, in the colors of Figure~\ref{fig:rulings}. In round 1 of the consultative run the chair asked the resident and the radiologist for opinions. These roles speak only in Stage I, so the request went unanswered and round 2 was again between the deputy and the rehabilitation physician. A ruling to work up first records in a structured field whether the workup would lead to surgery or to conservative care, and a ruling with two branches is coded by the branch the chair enters there. The round-2 ruling (Figure~\ref{fig:chat-consultative}, right) has two branches: work up first, operate if the new imaging shows cord signal change or instability, and otherwise start a conservative trial. In its structured output the chair recorded that the workup would lead to surgery, so the run counted as surgery intended (Section~\ref{sec:measures}).

\FloatBarrier

\section{Redaction and its verification}\label{app:redaction}

The redaction prompt is in the repository at \path{prompts/case_report_filter.md}, and the checker at \path{services/case_report_filter.py}. The output must be an in-order deletion of the source. A full stop, semicolon or comma may be added only where text was deleted, and no retained sentence may lose just one or two words, so that a deletion cannot reverse what a sentence says. On a violation the reason is attached to the next request, with at most three retries. All 20 cases passed at the first attempt. For each of the 20 cases, the redaction record (deleted fragments, categories of deletion and residual keyword hits) is saved with the run summary.

\section{Results by case}\label{app:cases}

Cases are numbered C1--C20 in the order of Figure~\ref{fig:rulings}. Entries are rulings to operate now out of runs.

\begin{center}
\scriptsize
\setlength{\tabcolsep}{1.5pt}
\begin{tabularx}{\textwidth}{@{}P{0.5cm}P{1.55cm}P{1.3cm}P{1.45cm}*{7}{C}@{}}
\toprule
 & & & & & \multicolumn{3}{c}{Rules version} & \multicolumn{3}{c}{Trait-only version} \\
\cmidrule(lr){6-8}\cmidrule(l){9-11}
No. & PMC-Patients (PMID) & Group & Actual management & Neutral & Consultative & Autocratic & Avoidant & Consultative & Autocratic & Avoidant \\
\midrule
C1 & 8059481-1 (33898127) & Borderline or not operated & Not operated & 1/3 & 1/3 & 3/3 & 0/3 & 1/2 & 2/2 & 0/2 \\
C2 & 6580147-1 (31207532) & Borderline or not operated & Delayed surgery & 3/3 & 1/3 & 3/3 & 1/3 & 1/2 & 2/2 & 1/2 \\
C3 & 6389466-1 (30847190) & Borderline or not operated & Delayed surgery & 3/3 & 2/3 & 3/3 & 0/3 & 1/2 & 2/2 & 1/2 \\
C4 & 7529124-1 (33071672) & Borderline or not operated & Surgery & 1/3 & 1/3 & 3/3 & 0/3 & 2/2 & 2/2 & 0/2 \\
C5 & 7227285-1 (32410709) & Borderline or not operated & Surgery & 1/3 & 0/3 & 3/3 & 0/3 & 0/2 & 1/2 & 0/2 \\
C6 & 2857503-2 (20411153) & Borderline or not operated & Surgery & 1/3 & 0/3 & 3/3 & 0/3 & 0/2 & 1/2 & 0/2 \\
C7 & 7440276-1 (32839680) & Difficult diagnosis & Surgery & 1/3 & 0/3 & 3/3 & 0/3 & 2/2 & 1/2 & 0/2 \\
C8 & 6538236-1 (31183272) & Difficult diagnosis & Surgery & 3/3 & 2/3 & 3/3 & 0/3 & 0/2 & 2/2 & 0/2 \\
C9 & 5553503-1 (28825009) & Difficult diagnosis & Surgery & 0/3 & 2/3 & 2/3 & 0/3 & 0/2 & 0/2 & 0/2 \\
C10 & 8645501-1 (34877036) & Difficult diagnosis & Surgery & 1/3 & 2/3 & 3/3 & 0/3 & 1/2 & 2/2 & 1/2 \\
C11 & 7989791-2 (33790773) & Difficult diagnosis & Surgery & 0/3 & 0/3 & 1/3 & 0/3 & 0/2 & 0/2 & 0/2 \\
C12 & 6187008-1 (30370039) & Difficult diagnosis & Surgery & 2/3 & 3/3 & 3/3 & 0/3 & 0/2 & 2/2 & 0/2 \\
C13 & 5482166-1 (28680730) & Difficult diagnosis & Surgery & 2/3 & 1/3 & 3/3 & 0/3 & 1/2 & 2/2 & 0/2 \\
C14 & 4426531-1 (25972718) & Difficult diagnosis & Surgery & 3/3 & 3/3 & 3/3 & 0/3 & 2/2 & 2/2 & 1/2 \\
C15 & 5159182-1 (27887013) & Clear indication & Surgery & 3/3 & 3/3 & 3/3 & 0/3 & 2/2 & 2/2 & 0/2 \\
C16 & 6254142-1 (30474608) & Clear indication & Surgery & 2/3 & 2/3 & 3/3 & 0/3 & 1/2 & 2/2 & 0/2 \\
C17 & 7933642-1 (33708711) & Clear indication & Surgery & 1/3 & 3/3 & 3/3 & 0/3 & 0/2 & 1/2 & 1/2 \\
C18 & 4432380-1 (25983784) & Clear indication & Surgery & 3/3 & 2/3 & 3/3 & 0/3 & 1/2 & 2/2 & 1/2 \\
C19 & 6533938-1 (31124460) & Clear indication & Surgery & 3/3 & 2/3 & 3/3 & 0/3 & 1/2 & 2/2 & 0/2 \\
C20 & 2729274-1 (19718247) & Clear indication & Surgery & 0/3 & 1/3 & 2/3 & 0/3 & 0/2 & 1/2 & 0/2 \\
\bottomrule
\end{tabularx}
\end{center}

\section{The red-flag judge}\label{app:judge}

The judge prompt is in the repository at \path{prompts/flag_handling_judge.md}. The judge read only the neurologist's list of flags and the chair's rulings, not the arm, the persona prefixes or the statements of other roles. The agreement check sampled 40 runs and 124 flags: per-flag agreement was 91.1\%, and Cohen's $\kappa = 0.836$.

\section{Statistical details}\label{app:stats}

\subsection{Tests}

The primary analysis paired by case. For each case and arm we took the share of runs ruled to operate now and subtracted the neutral arm's share, giving 20 differences. Exact sign-flip permutation tests over all sign combinations were one-sided for H1 (autocratic above neutral) and H2 (avoidant below neutral) and two-sided for H3 (consultative different from neutral), with Holm correction across the three contrasts. The 95\% CIs for the mean difference came from a case-level bootstrap with 10,000 resamples. H6a--c, labeled H1t--H3t in the frozen plan, repeated these tests on the trait-only arms, with the neutral runs of the rules version as control. For H4 we took, per case, the absolute difference from neutral in operating now minus the absolute difference in surgery intended, and bootstrapped the mean of this gap over cases. H5 was descriptive. Secondary endpoints used the same tests with BH control of the false discovery rate. H7a--c used the same tests on the share of decision-bearing flags not addressed (H7a, H7c) or acted on (H7b) in the 9 cases that had such flags, with Holm correction across the three, separately for the rules and the trait-only versions. All flags together were analyzed as exploratory.

\subsection{Sensitivity analyses (rules version, against neutral)}

Each sensitivity analysis checks one alternative reading of the primary test. The random-intercept logistic regression and the generalized estimating equations (GEE, exchangeable correlation) model each run instead of each case's share of runs, to check that the effects do not depend on the paired test. The regression was fitted by variational Bayes (statsmodels' \texttt{BinomialBayesMixedGLM}) and reports posterior means and standard deviations instead of $p$ values. The McNemar test on each cell's majority ruling checks that the effects hold when each cell contributes a single ruling. Dropping the 11 cases whose redacted records still contained keyword hits related to the decision (all read by the authors and found to be past history or tests already done) checks that the effects do not come from answers leaked through the redaction.

\begin{center}
\footnotesize
\begin{tabularx}{\textwidth}{@{}L*{3}{Q{2.4cm}}@{}}
\toprule
Method & Autocratic & Avoidant & Consultative \\
\midrule
Random-intercept logistic regression (variational Bayes; log odds ratio, posterior SD) & 2.92 (0.51) & $-4.46$ (0.69) & $-0.22$ (0.29) \\
GEE, exchangeable correlation (log odds ratio, $p$) & 2.37 ($1\times10^{-6}$) & $-4.35$ ($6\times10^{-6}$) & $-0.20$ (0.53) \\
McNemar test on majority rulings (only the persona arm operates : only neutral operates, $p$) & 9:0 (0.004) & 0:10 (0.002) & 3:2 (1) \\
Without the 11 cases with residual hits (9 cases; difference in points, Holm $p$) & +33 (0.06) & $-56$ (0.02) & 0 (1) \\
\bottomrule
\end{tabularx}
\end{center}

\subsection{Chance-corrected agreement (exploratory)}

This analysis was added after the plan was frozen. For each arm of the rules version, the 20 cases are the subjects and the three runs the raters, over five ruling categories (operate now, work up first, conservative care, postpone, not yet decided). The full agreement expected by chance is $\sum_k p_k^3$, where $p_k$ is the arm's share of rulings in category $k$. Fleiss' $\kappa$ \cite{Fleiss1971} was bootstrapped over cases with the seed and the 10,000 resamples of the main analysis. A resample in which every ruling falls in one category leaves $\kappa$ undefined and was dropped: 387 resamples for the autocratic and 1,233 for the avoidant arm. The difference from neutral resampled the same cases for both arms. The Spearman correlation between an arm's and the neutral arm's per-case shares of rulings to operate now was tested with 100,000 permutations.

\begin{center}
\footnotesize
\setlength{\tabcolsep}{3.5pt}
\begin{tabularx}{\textwidth}{@{}P{2.3cm}*{5}{C}@{}}
\toprule
Arm (rules version) & All three runs agree & Expected by chance & Fleiss $\kappa$ (95\% CI) & $\kappa$ minus neutral (95\% CI) & Spearman $\rho$ with neutral ($p$) \\
\midrule
Neutral & 50\% & 25\% & 0.31 (0.02, 0.58) & -- & -- \\
Consultative & 40\% & 25\% & 0.20 ($-0.10$, 0.47) & $-0.11$ ($-0.54$, 0.33) & 0.50 (0.03) \\
Autocratic & 85\% & 81\% & 0.20 ($-0.07$, 0.48) & $-0.11$ ($-0.54$, 0.34) & 0.65 ($<$0.001) \\
Avoidant & 90\% & 86\% & 0.31 ($-0.05$, 0.48) & 0.00 ($-0.56$, 0.42) & 0.27 (0.50) \\
\bottomrule
\end{tabularx}
\end{center}

\subsection{Secondary endpoints (difference from neutral, BH-corrected $p$)}

\begin{center}
\footnotesize
\setlength{\tabcolsep}{3.5pt}
\begin{tabularx}{\textwidth}{@{}P{3.1cm}*{6}{C}@{}}
\toprule
 & \multicolumn{3}{c}{Rules version} & \multicolumn{3}{c}{Trait-only version} \\
\cmidrule(lr){2-4}\cmidrule(l){5-7}
Endpoint & Consultative & Autocratic & Avoidant & Consultative & Autocratic & Avoidant \\
\midrule
Surgery judged indicated (points) & $-3$ (1) & +5 (0.68) & $-40$ (0.0002) & $-3$ (1) & $-1$ (1) & $-21$ (0.07) \\
Work up first (points) & +7 (0.70) & $-35$ (0.0006) & +53 (0.0001) & +18 (0.11) & $-19$ (0.015) & +41 (0.004) \\
Family involvement (points) & 0 (1) & 0 (1) & +78 ($<$0.0001) & +3 (1) & 0 (1) & +8 (0.75) \\
Deputy objections (per run) & +2.8 ($<$0.0001) & $-2.4$ ($<$0.0001) & +0.8 (0.017) & +1.3 (0.004) & $-0.4$ (0.12) & +1.4 (0.004) \\
\bottomrule
\end{tabularx}
\end{center}

\subsection{Handling of red flags (difference from neutral in percentage points; Holm-corrected $p$ within each set of three tests)}

\begin{center}
\footnotesize
\begin{tabularx}{\textwidth}{@{}L*{2}{Q{3.2cm}}@{}}
\toprule
 & Rules version & Trait-only version \\
\midrule
Decision-bearing flags, autocratic, not addressed (H7a, 9 cases) & +7 (0.75) & +15 (0.75) \\
Decision-bearing flags, avoidant, acted on (H7b) & $-4$ (1) & +1 (1) \\
Decision-bearing flags, consultative, not addressed (H7c) & +6 (1) & +1 (1) \\
All flags, autocratic, not addressed (exploratory, 20 cases) & +22 ($<$0.001) & +0.1 (0.71) \\
All flags, avoidant, acted on & +12 (0.012) & +11 (0.18) \\
All flags, consultative, not addressed & $-8$ (0.06) & $-9$ (0.71) \\
\bottomrule
\end{tabularx}
\end{center}

\section{Process measures by arm}\label{app:process}

\begin{center}
\footnotesize
\setlength{\tabcolsep}{3.5pt}
\begin{tabularx}{\textwidth}{@{}P{2.9cm}*{7}{C}@{}}
\toprule
\armheader
\midrule
Mean negotiation rounds & 1.07 & 1.85 & 1.00 & 1.60 & 1.45 & 1.08 & 1.48 \\
Ruled in the first round & 93\% & 15\% & 100\% & 40\% & 55\% & 92.5\% & 52.5\% \\
Tests ordered & 2.78 & 4.00 & 0.73 & 5.10 & 3.90 & 2.43 & 4.15 \\
Roles addressed by name & 5.23 & 3.88 & 4.30 & 5.47 & 5.23 & 5.08 & 4.65 \\
Plan has a contingency & 98\% & 100\% & 40\% & 100\% & 100\% & 100\% & 100\% \\
Family involved in deciding & 0 & 0 & 0 & 78\% & 2.5\% & 0 & 7.5\% \\
Deputy objections (per run) & 3.40 & 6.17 & 1.00 & 4.17 & 4.73 & 3.03 & 4.80 \\
Deputy objections (first turn) & 3.15 & 3.42 & 1.00 & 3.05 & 3.40 & 2.93 & 3.38 \\
\bottomrule
\end{tabularx}
\end{center}

\end{document}